\documentclass{sf2a-conf2022}
\usepackage{graphicx}
\usepackage{hyperref}
\usepackage[]{natbib}  
\usepackage{epstopdf}

\def\BibTeX{{\rm B\kern-.05em{\sc i\kern-.025em b}\kern-.08em
    T\kern-.1667em\lower.7ex\hbox{E}\kern-.125emX}}
\bibpunct{(}{)}{;}{a}{}{,}  %%%%%%%%%%%%%  A&A bibliography style
\usepackage{eurosym}
\newcommand{\celsius}{\mbox{$^\circ$C}}

\newcommand{\meuro}{\mbox{M\euro}}
\newcommand{\co}{\mbox{CO$_2$}}
\newcommand{\coe}{\mbox{CO$_2$e}}
\newcommand{\tco}{\mbox{tCO$_2$}}

\newcommand{\tcoe}{\mbox{tCO$_2$e}}
\newcommand{\ktcoe}{\mbox{ktCO$_2$e}}

\newcommand{\Mtcoe}{\mbox{MtCO$_2$e}}
\newcommand{\tcoeyr}{\mbox{tCO$_2$e yr$^{-1}$}}
\newcommand{\ktcoeyr}{\mbox{ktCO$_2$e yr$^{-1}$}}
\newcommand{\Mtcoeyr}{\mbox{MtCO$_2$e yr$^{-1}$}}
\newcommand{\moneyeft}{\mbox{tCO$_2$e \meuro$^{-1}$}}
\newcommand{\masseft}{\mbox{tCO$_2$e kg$^{-1}$}}

\begin{document}

\TitreGlobal{SF2A 2022}

%%%%%%%%%%%%%%%%%%%%%%%%%%%%%%%%%%%%%%%%%%%%%%%%%%
% Titlepage
%%%%%%%%%%%%%%%%%%%%%%%%%%%%%%%%%%%%%%%%%%%%%%%%%%
\title{The carbon footprint of astronomical research infrastructures}

\runningtitle{The carbon footprint of astronomical research infrastructures}

\author{J. Kn\"odlseder}\address{Institut de Recherche en Astrophysique et Plan\'etologie, Universit\'e de Toulouse, CNRS, CNES, 
9 avenue Colonel Roche, 31028 Toulouse, Cedex 4, France}

%% Keep this line, even if the page will be settled afterwards.
\setcounter{page}{237}

%%-----------------------------------------------------------------

\maketitle

%%-----------------------------------------------------------------
%%        The abstract
%% 
%%  Warning!  within the abstract:
%%  - do not use macros. 
%%  - do not use commands like: \cite, \citet, \citep ... etc.

\begin{abstract}
We estimate the carbon footprint of astronomical research infrastructures, including space telescopes 
and probes and ground-based observatories.
Our analysis suggests annual greenhouse gas emissions of $1.2\pm0.2$ \Mtcoeyr\ due to construction
and operation of the world-fleet of astronomical observatories, corresponding to a carbon footprint of
36.6$\pm$14.0 \tcoe\ per year and average astronomer. 
We show that decarbonising astronomical facilities is compromised by the continuous deployment
of new facilities, suggesting that a significant reduction in the deployment pace of new facilities is needed 
to reduce the carbon footprint of astronomy.
We propose measures that would bring astronomical activities more in line with the imperative
to reduce the carbon footprint of all human activities.
\end{abstract}

%% Insert the keywords (to appear in the ADS indexing)
%% Keywords must be separated by a comma

%%-----------------------------------------------------------------

%%%%%%%%%%%%%%%%%%%%%%%%%%%%%%%%%%%%%%%%%%%%%%%%%%
% Introduction
%%%%%%%%%%%%%%%%%%%%%%%%%%%%%%%%%%%%%%%%%%%%%%%%%%
\section{Introduction}

The Intergovernmental Panel on Climate Change (IPCC) is a body created in 1988 by the World 
Meteorological Organization (WMO) and the United Nations Environment Programme (UNEP)
with the objective to provide governments at all levels with scientific information that they can 
use to develop climate policies.
The IPCC authors assess thousands of scientific papers published each year to provide a 
comprehensive summary of what is known about the drivers of climate change, its impacts and 
future risks, and how adaptation and mitigation can reduce those risks.
According to the 6$^{\rm th}$ IPCC assessment report \citep{ipcc2021wg1}, it is unequivocal that human influence 
has warmed the atmosphere, ocean and land.
The scale of recent changes across the climate system as a whole -- and the present state of many
aspects of the climate system -- are unprecedented over many centuries to many thousands of years.
Global warming of 1.5\celsius\ and 2\celsius\ will be exceeded during the 21$^{\rm st}$ century unless 
deep reductions in carbon dioxide (\co) and other greenhouse gas (GHG) emissions occur in the coming 
decades.
Many changes due to past and future GHG emissions are irreversible for centuries to 
millennia, especially changes in the ocean, ice sheets and global sea level.
From a physical science perspective, limiting human-induced global warming to a specific level 
requires limiting cumulative \co\ emissions, reaching at least net zero \co\ emissions, along with 
strong reductions in other GHG emissions.

There is growing recognition in the astronomy and astrophysics community that it must assume its 
share in the global effort to reduce GHGs.
Like many other institutes we have therefore undertaken at the Institut de Recherche en Astrophysique 
et Plan\'etologie (IRAP) an estimate of our GHG emissions so that we can devise an action plan 
that meets the challenge to drastically reduce emissions.
In doing this exercise, we aimed in including all relevant sources of GHG emissions, comprising the
purchase of goods and services and the use of data from research infrastructures, such as space 
telescopes and probes and ground-based observatories.
While these sources were generally omitted in other works, the Bilan Carbone\textsuperscript{\textcopyright} 
method that we used for our estimate prescribes to include all sources for which our laboratory is 
responsible and on which our activity depends on.
In other words, to identify the sources that need to be included in the estimate, the question to ask
is whether our activity will be impacted if a given source is removed.
Obviously, removing purchase of goods and services and use of data from observing facilities would
make our activity impossible.
In addition, research infrastructures are often invented, eventually built, and sometimes operated by 
researchers from our lab, hence as astronomers we also share the responsibility for their existence.

In total we found that IRAP's GHG emissions in 2019 were $51.5\pm6.0$ \tcoe\ per 
astronomer of which $27.4\pm4.8$ \tcoe\ were attributed to the use of observational data
\citep{martin2022}.
Interestingly, the sources that were so far neglected in GHG emission estimates of other research
laboratories dominate IRAP's GHG emissions, with 55\% due to the use of observational data and 
18\% due to the purchase of goods and services, of which a substantial fraction is related to instrument 
developments.
The next most important source of GHG emissions was professional travelling (16\%), all
remaining contributions sum up to only 11\%.
So IRAP's carbon footprint is largely dominated by inventing, developing, constructing and using 
research infrastructures, which is the core business of our institute.
To understand whether this is specific to IRAP, or whether this is a general feature of astronomy, 
we went one step further and estimated the total carbon footprint of the world-fleet of astronomical 
observatories that were operating in 2019.

%%%%%%%%%%%%%%%%%%%%%%%%%%%%%%%%%%%%%%%%%%%%%%%%%%
% Estimate of the carbon footprint of the world-fleet of astronomical research infrastructures
%%%%%%%%%%%%%%%%%%%%%%%%%%%%%%%%%%%%%%%%%%%%%%%%%%
\section{Estimate of the carbon footprint of the world-fleet of astronomical research infrastructures}

%%%%%%%%%%%%%%%%%%%%%%%%%%%%%%%%%%%%%%%%%%%%%%%%%%
\subsection{Method}

We estimated the carbon footprint of astronomical research infrastructures using primarily a
monetary method that relates cost to GHG emissions.
This approach is known to have large uncertainties due to the aggregation of activities, products and 
monetary flows that may vary considerably from one facility or field of activity to another. 
An alternative life-cycle assessment (LCA) methodology is recommended by key space industry 
actors \citep{esahandbook} as the optimal method to assess and reduce the carbon footprint 
of space missions, but it is difficult to implement in practice (especially for comparative or discipline-wide 
assessments) due to the confidential nature of the required input activity data \citep{maury2020}.
At present, a monetary method analysis is thus the only feasible way to assess the combined carbon 
footprint of the world's space- and ground-based astronomical research infrastructures.
For space missions, we complemented the monetary method by an alternative approach based on
the payload launch mass.
We adopted throughout this study an uncertainty of 80\% for the carbon footprint estimate of individual 
facilities, as recommended by the French Agency for Ecological Transition (ADEME) for a monetary
analysis \citep{basecarbone}.

For our estimate we followed the standard method of multiplying activity data with emission factors, 
including GHG emissions from constructing and operating the facilities.
We started with considering a list of facilities from which data were used in peer-reviewed journal 
articles made by IRAP researchers in 2019.
The list includes 46 space missions and 39 ground-based observatories.
For space missions, we estimated the carbon footprint by multiplying mission cost or payload launch 
mass with appropriate emission factors.
Owing to their longer lifetimes compared to space missions, we separated construction from
operations for ground-based observatories and estimated the carbon footprint by multiplying 
construction and operating costs with appropriate emission factors.
The full list of cost and mass data that we gathered from the literature and the internet can be
found in the Supplementary Information of \citet{knodlseder2022}.

To derive the carbon footprint of the world-fleet of astronomical facilities we only considered 
the infrastructures that were still operating in 2019, reducing our initial list from 85 to 75 
facilities.
We then used a bootstrap method to extrapolate the carbon footprint of the facilities in our
list to an estimated number of 55 active space missions and 1142 ground-based observatories.\footnote{We
estimated that there are $\sim1000$ optical or near infrared telescopes with diameters
smaller than 3 metres in the world, which largely dominates the number of observatories
but provides only a modest contribution to the aggregated carbon footprint.}
In short, the bootstrap method randomly selects $M$ facilities from a reduced list of $N$
infrastructures, selecting on average each infrastructure $M/N$ times (with $M \geq N$).
Summing up the carbon footprints of all selected infrastructures provides then a linear 
extrapolation of the carbon footprint from $N$ to $M$ infrastructures.
Yet bootstrapping goes beyond a linear extrapolation in that it preserves the discrete character
of the facilities, and by repeating the sampling process it provides a probability density distribution
for the aggregated carbon footprint of the $M$ facilities.
We repeated the random sampling 10,000 times and used the mean and standard deviation 
of the results to provide an estimate of the value and uncertainty of the aggregated carbon footprint.
In order to reduce the bias that may arise from the specific 75 facilities in our initial list, and
to avoid mixing infrastructures with hugely different carbon footprints (such as small and large
optical telescopes), we divided the facilities in our list into broad categories that reflect scientific 
topic and observatory type.
Details of the method and estimates for the number of worldwide active facilities per category are
provided in \citet{knodlseder2022}.

%%%%%%%%%%%%%%%%%%%%%%%%%%%%%%%%%%%%%%%%%%%%%%%%%%
\subsection{Emission factors}

We estimated dedicated emission factors for our study using existing carbon footprint estimates
for space missions and ground-based observatories.
Specifically, life-cycle carbon footprints of space missions were estimated from the case studies
of \citet{wilson2019} which covered the entire mission including the launcher and a few years of 
operations.
From these studies, we infered mean emission factors of 140 \tco\ equivalent (\coe) per million \euro\ 
(\meuro) of mission cost and 50 \masseft\ of payload launch mass. 
Emission factors of ground-based observatories were derived using existing carbon footprint 
assessments for the construction of two facilities and the operations of three facilities. 
We found a mean emission factor of 240 \moneyeft\ for construction and of 250 \moneyeft\ for operations.
A lower monetary emission factor for space missions is supported by the fact that space missions 
are much less material intensive compared with ground-based observatories after normalizing by cost. 
For example, the liftoff mass of a \euro 1 billion space mission launched with Ariane 5 ECA is about 
790 tonnes, while the European Extremely Large Telescope (E-ELT), which has a similar cost, has a 
mass of about 60,000 tonnes. 
The space sector is in fact unique, and is characterized by low production rates, long development 
cycles and specialized materials and processes \citep{geerken2018}.

%%%%%%%%%%%%%%%%%%%%%%%%%%%%%%%%%%%%%%%%%%%%%%%%%%
\begin{table}[t!]
\caption{Emission factors.
\label{tab:emission-factors}}
\centering
\begin{tabular}{l c}
\hline
Activity & Emission factor \\
\hline
Space missions (based on payload launch mass) & 50 \masseft\ \\
Space missions (based on mission cost) & 140 \moneyeft\ \\
Ground-based observatory construction & 240 \moneyeft\ \\
Ground-based observatory operations & 250 \moneyeft\ \\
\hline
Insurance, banking and advisory services & 110 \moneyeft\ \\
Architecture and engineering, building maintenance & 170 \moneyeft\ \\
Installation and repair of machines and equipment & 390 \moneyeft\ \\
Metal products (aluminum, cupper, steel, etc.) & 1700 \moneyeft\ \\
Mineral products (concrete, glass, etc.) & 1800 \moneyeft\ \\
\hline
\end{tabular}
\end{table}
%%%%%%%%%%%%%%%%%%%%%%%%%%%%%%%%%%%%%%%%%%%%%%%%%%

The emission factors used in this study are summarised in Table \ref{tab:emission-factors} where
they are compared to monetary emission factors selected from \citet{basecarbone}, covering 
the range of values encountered for economic activity sectors in France.
The comparison shows that emission factors for astronomical research infrastructures are at the
low side of other economic activity sectors, implying that decarbonising observatory construction and
operations will be challenging within the current socio-economic system.
Office work is an important contributor to the carbon footprint of the space sector \citep{chanoine2017},
which is in agreement with the observation that its emission factor is close to that of office
work activities such as insurance, banking and advisory services.
As explained above, constructing ground-based observatories is considerably more material intensive 
than building a space mission, hence a larger emission factor for ground-based observatory construction
with respect to space missions is plausible.

Due to the lack of published information we were not able to derive a specific emission factor for 
space mission operations, yet since the underlying infrastructures and activities are similar to
operations of ground-based observatories it seems plausible that their emission factors are
comparable.
We note that the emission factor of operations depends sensitively on the carbon intensity of electricity 
generation (which is an important contributor to the overall operations footprint) and the number
of persons needed for operations (which is an important contributor to the overall operating costs).
Consequently, the operations emission factor for a specific facility may deviate significantly from our 
estimated average value, yet since we are considering here only the aggregated carbon footprint of 
astronomical facilities such deviations should average out.

%%%%%%%%%%%%%%%%%%%%%%%%%%%%%%%%%%%%%%%%%%%%%%%%%%
\subsection{Results}

The aggregated results of our estimation are summarised in Table \ref{tab:carbonfootprint}.
Two set of values are presented: the first where we bootstrap-sampled all research infrastructures
in each of the categories, and the second where we bootstrap-sampled all except the facilities
with the largest carbon footprint in each category (the footprints of the non-sampled facilities
were simply added to the bootstrap result).
The latter approach is motivated by the possibility that the largest carbon footprint in a given
category arises from a facility that is unique in the world, hence excluding this facility from the
sampling avoids that the bootstrap sampling selects this unique facility multiple times.
Examples for such unique facilities in our initial list are the Hubble space telescope or the
ALMA observatory which have annual carbon footprints of several tens of \ktcoeyr.
So the second approach is more conservative, plausibly bracketing together with the first approach
the true value of the carbon footprint of astronomical research infrastructures.

%%%%%%%%%%%%%%%%%%%%%%%%%%%%%%%%%%%%%%%%%%%%%%%%%%
\begin{table}[t!]
\caption{Carbon footprint of world-fleet of astronomical research infrastructures active in 2019.
\label{tab:carbonfootprint}}
\centering
\begin{tabular}{l c c}
\hline
Category & Lifecycle footprint (\Mtcoe) & Annual footprint (\ktcoeyr) \\
\hline
\multicolumn{3}{c}{All facilities sampled} \\
Space missions (cost-based) & $8.4\pm2.0$ & $596\pm111$ \\
Space missions (mass-based) & $6.4\pm1.2$ & $455\pm74$ \\
Ground-based observatories & $14.2\pm1.5$ & $757\pm131$ \\
Total & $21.6\pm3.2$ & $1283\pm232$ \\
\hline
\multicolumn{3}{c}{Facility with largest footprint excluded from sampling} \\
Space missions (cost-based) & $7.1\pm1.4$ & $490\pm79$ \\
Space missions (mass-based) & $5.8\pm0.9$ & $417\pm65$ \\
Ground-based observatories & $12.6\pm1.0$ & $600\pm70$ \\
Total & $19.0\pm2.3$ & $1054\pm137$ \\
\hline
Total (average) & $20.3\pm3.3$ & $1168\pm249$ \\
\hline
\end{tabular}
\end{table}
%%%%%%%%%%%%%%%%%%%%%%%%%%%%%%%%%%%%%%%%%%%%%%%%%%

Table \ref{tab:carbonfootprint} gives both the lifecycle and the annual footprints.
The lifecycle footprint includes the contributions from construction and operations until 2019,
while the annual footprint is the sum of the lifecycle footprint of each facility divided by its
lifetime, defined as the time since start of operations, or ten years, whatever is longer.
While the lifecycle footprint aggregates carbon footprints over different time periods, and hence
is of limited use, the annual footprint is an estimate of the yearly GHG emissions of the
considered research infrastructures.
The last row provides the average results between both bootstrapping approaches, with the 
differences between the results added to the quoted uncertainties.
Our analysis hence suggests that the world-fleet of astronomical facilities that were operating 
in 2019 had an annual carbon footprint of $1.2\pm0.2$ \Mtcoeyr.

Dividing the annual carbon footprint by an estimated number of 30,000 astronomers in the world
gives a footprint of $42.8\pm7.7$ \tcoeyr\ per average astronomer for the first bootstrapping
approach and $35.1\pm4.6$ \tcoeyr\ for the second.
These results are a bit larger than the estimated footprint of $27.4\pm4.8$ \tcoeyr\ related to the 
use of observational data for an average IRAP astronomer, yet the IRAP estimate is based on a
restricted list of facilities which may tend to underestimate the true footprint.
Taking nevertheless all these results at face value, we derive an estimate of 
$36.6\pm14.0$ \tcoeyr\ for the annual carbon footprint of the world-fleet of astronomical facilities
per average astronomer that comprises all individual results and their uncertainties.

%%%%%%%%%%%%%%%%%%%%%%%%%%%%%%%%%%%%%%%%%%%%%%%%%%
% Consequences
%%%%%%%%%%%%%%%%%%%%%%%%%%%%%%%%%%%%%%%%%%%%%%%%%%
\section{Consequences}

According to our analysis, astronomical research infrastructures appear to be the single most 
important contributor to the carbon footprint of an average astronomer.
Additional contributions include purchase of goods and services, travelling and commuting, 
supercomputing, running the office building and meals, that for IRAP add up to an
additional carbon footprint of $23$ \tcoeyr\ per astronomer, resulting in a total professional
annual footprint of an average astronomer of about $\sim50$ \tcoeyr.
Adding also the astronomer's lifestyle footprint, estimated to $10$ \tcoeyr\ for upper class 
consumers in France \citep{lenglart2010}, leads to an estimated annual footprint of about 
$\sim60$ \tcoeyr\ for an average astronomer in France.

Keeping global warming with a reasonable chance below a level of 1.5$\,^{\circ}$C or 2$\,^{\circ}$C
requires GHG emission reductions by 84\% or 63\% in 2050 with respect to 2019 \citep[][]{ipcc2022wg3}, 
corresponding to annual average emission reductions of about $\sim6\%$ or $\sim3\%$.
GHG emissions are not equally distributed between regions, activities and humans, requiring more 
than average reductions by important emitters to assure the social acceptability of the
efforts.
Our analysis suggests that astronomers are important emitters, and asking consequently for an order 
of magnitude reduction effort of GHG emissions over the coming 30 years is not implausible.
Obviously, astronomy has not only an environmental but also a societal impact, and finding the 
right balance between these impacts needs to be subject of public debate.
Yet this applies to all sectors of human activity, be it any scientific sector, or sectors that satisfy 
basic human needs, such as agriculture, housing, health care, dressing and transport.
Exempting astronomy from significant GHG emission reductions seems thus difficult to justify.

%%%%%%%%%%%%%%%%%%%%%%%%%%%%%%%%%%%%%%%%%%%%%%%%%%
% Taking action
%%%%%%%%%%%%%%%%%%%%%%%%%%%%%%%%%%%%%%%%%%%%%%%%%%
\section{Taking action}

Coming back to astronomical research infrastructures, reducing their carbon footprint requires
first that each planned or existing facility performs a detailed environmental lifecycle analysis,
informing quantified action plans to reduce their emissions.
Progress in the implementation of the action plans need to be monitored, and plans be adapted
if needed.
LCA results, action plans and achievements need to be made public, so that the progress on GHG
emission reductions is transparent and fairness can be assured.

%%%%%%%%%%%%%%%%%%%%%%%%%%%%%%%%%%%%%%%%%%%%%%%%%%
\begin{figure}[t!]
 \centering
 \includegraphics[width=0.8\textwidth,clip]{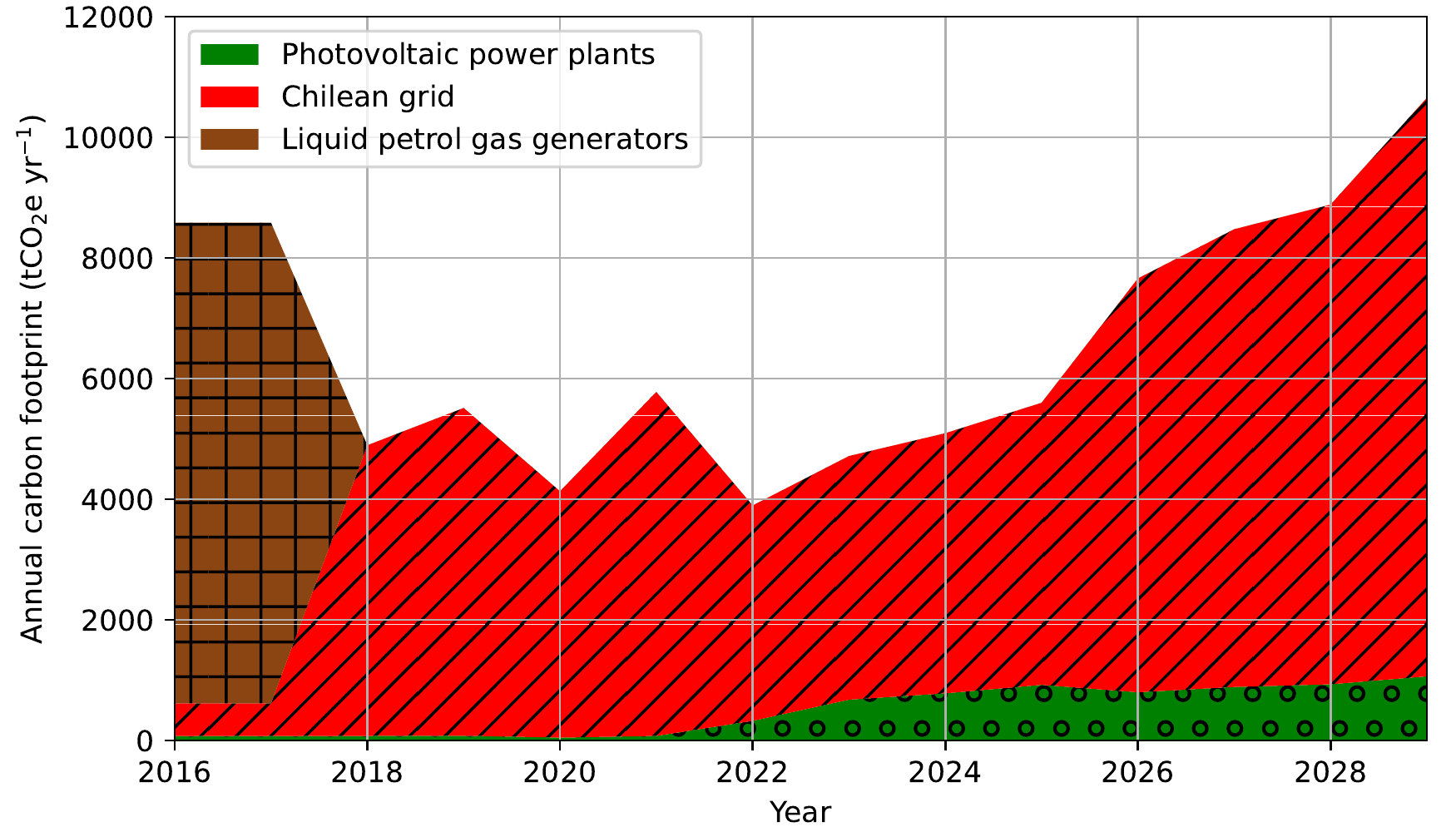}      
  \caption{Past and predicted annual carbon footprint of electricity consumption at the ESO 
  observatory sites in La Silla, Paranal and Armazones \citep[data from][]{filippi2022}.}
  \label{fig:esofootprint}
\end{figure}
%%%%%%%%%%%%%%%%%%%%%%%%%%%%%%%%%%%%%%%%%%%%%%%%%%

For proposed facilities LCA results should inform implementation decisions, while for existing
facilities LCA results should inform decarbonisation plans.
Possible actions include switching to renewable energies for observatory operations, reducing 
air-travelling and avoiding air-shipping, moving to electric vehicle fleets, and extending  
equipment lifetime.
With such measures, the European Southern Observatory (ESO) plans to reduce its 
operations-related GHG emissions of 28 \ktcoeyr\ in 2018 by up to 4.4 \ktcoeyr\ over the next years,
corresponding to a reduction of 15\%.\footnote{\url{https://www.eso.org/public/about-eso/green/}}
This is an important step, yet falls short of the required reduction levels mentioned above.
In addition, ESO is currently building the E-ELT with an
estimated construction carbon footprint of at least 63.7 \ktcoe\ (ESO, personal communication), 
corresponding to about 15 years of GHG emission savings.
Operating the E-ELT will add additional GHG emissions, as illustrated by the past and predicted annual 
carbon footprint of electricity consumption at the ESO observatory sites in Chile, shown in
Fig.~\ref{fig:esofootprint}.
While between 2016 and 2022 a reduction of GHG emissions from electricity consumption by $\sim50\%$ was 
achieved (by swapping at Paranal from liquid petrol gas generators to a grid connection in 2018 and 
adding photovoltaic power plants in 2022), the additional electricity needs of E-ELT will have annihilated 
all the reductions by the end of this decade; despite important efforts, the GHG emissions due to 
electricity consumption will exceed in 2030 those of 2016.
This illustrates an obvious but inconvenient truth: it is extremely difficult to decarbonise while ramping
up!
ESO is so far the only organisation that provides public information on carbon footprint estimates and 
reduction plans, exposing the organisation obviously to be used as a case-study.
There are no reasons to believe that the situation is different for other organisations, at least as long as 
they continue to expand.
It's up to these other organisations to prove us wrong, yet until this is done, we should accept that
reducing the GHG emissions of astronomy is challenging while continuing with the deployment of new
facilities at the current pace.

%%%%%%%%%%%%%%%%%%%%%%%%%%%%%%%%%%%%%%%%%%%%%%%%%%
% Towards sustainable astronomy
%%%%%%%%%%%%%%%%%%%%%%%%%%%%%%%%%%%%%%%%%%%%%%%%%%
\section{Towards sustainable astronomy}

Obviously, all this calls for deep changes in how astronomy is done in the future, but given the required 
order of magnitude reductions in GHG emissions, how could it be otherwise?
A first step would be to use what we already have and move towards a more extended and deeper 
analysis of existing astronomical data archives.
It is well recognised that archives are valuable resources for astronomy, and a significant
fraction of discoveries is made by exploring already existing data \citep[e.g.][]{white2009,demarchi2022}.
Use of archival data should be actively promoted and be considered when evaluating operation
extensions.
Resources should be allocated according to carbon footprint, having in mind that remaining carbon
budgets that keep global warming below 1.5$\,^{\circ}$C or 2$\,^{\circ}$C shrink rapidly.
Today, no funding agency is investing significantly into decarbonising research infrastructures;
tomorrow, decarbonising existing facilities must become their funding priority!
This also means that less money will be available to build new infrastructures, yet is this really
a problem?
\citet{stoehr2015} argue that, in the future, observatories will compete for astronomers to work with 
their data, which if true seems to indicate that we may have already too many facilities.
There is no requirement {\em per se} on the deployment pace of new facilities or missions, and
slowing down the current pace will lead to less GHG emissions, free resources for investing into 
decarbonisation and give more time for in-depth analyses of existing data.
Another measure is moving away from competition towards more collaboration.
If we really believe that astronomers are working for mankind, there is no need to build the same kind
of facility several times on the globe.
For example, one 40-m class telescope in the world should be sufficient to make the discoveries to 
be made with such an instrument.
And there is no scientific justification for having a new space-race towards the planets, a few well-coordinated
international missions should be sufficient to gain the knowledge we are after.
Of course, astronomy is not the root cause of climate change, nor can astronomy alone fix it,
but astronomy with its significant per capita GHG emissions must be exemplary and take its fair share,
leading the way towards a sustainable future on Earth.

%%%%%%%%%%%%%%%%%%%%%%%%%%%%%%%%%%%%%%%%%%%%%%%%%%
% Acknowledgements
%%%%%%%%%%%%%%%%%%%%%%%%%%%%%%%%%%%%%%%%%%%%%%%%%%
\begin{acknowledgements}
The author would like to thank 
R. Arsenault,
S. Brau-Nogu{\'e},
M. Coriat,
P. Garnier,
A. Hughes,
P. Martin and
L. Tibaldo for useful discussions.
This work has benefited from discussions within the GDR Labos~1point5.
\end{acknowledgements}

\bibliographystyle{aa}  % A&A bibliography style file (aa.bst)
\bibliography{knodlseder} % your references in file: Yourfile.bib

\end{document}